%% file: main.tex
\documentclass[%
  a4paper,
  amsfonts,amssymb,amsmath,
  reprint,
  superscriptaddress,        
  showkeys,nofootinbib,twoside
]{revtex4-2}
\usepackage[english]{babel}
\usepackage[utf8]{inputenc}
\usepackage{amsthm}
\usepackage{siunitx}
\usepackage{comment}
\usepackage{graphicx}
\usepackage{subcaption}
\usepackage[colorinlistoftodos, color=green!40, prependcaption]{todonotes}
\usepackage{lineno}
\input{preamble}
\usepackage[pdftex, pdftitle={Article}, pdfauthor={Author}]{hyperref} 
\begin{document}
\title{Optically Active Single Hole Spin in ZnSe}

\author{Amirehsan Alizadehherfati}
\email{herfati@umd.edu}
\affiliation{Institute for Research in Electronics and Applied Physics and Joint Quantum Institute, University of Maryland, College Park, Maryland 20742, USA}
\affiliation{Department of Electrical and Computer Engineering, University of Maryland, College Park, MD 20742, USA}

\author{Yuxi Jiang}
\affiliation{Institute for Research in Electronics and Applied Physics and Joint Quantum Institute, University of Maryland, College Park, Maryland 20742, USA}
\affiliation{Department of Electrical and Computer Engineering, University of Maryland, College Park, MD 20742, USA}

\author{Kelsey J. Mirrielees}
\affiliation{Department of Materials Science and Engineering, North Carolina State University, Raleigh, North Carolina 27695, USA}

\author{Nils von den Driesch}
\affiliation{Peter-Gr\"unberg Institute (PGI-9 \& PGI-10), Forschungszentrum J\"ulich GmbH, 52428 J\"ulich, Germany}
\affiliation{JARA-Fundamentals of Future Information Technology, Forschungszentrum J\"ulich and RWTH Aachen University, 52062 Aachen, Germany}

\author{Christine Falter}
\affiliation{Peter-Gr\"unberg Institute (PGI-9 \& PGI-10), Forschungszentrum J\"ulich GmbH, 52428 J\"ulich, Germany}
\affiliation{JARA-Fundamentals of Future Information Technology, Forschungszentrum J\"ulich and RWTH Aachen University, 52062 Aachen, Germany}

\author{Yurii Kutovyi}
\affiliation{Peter-Gr\"unberg Institute (PGI-9 \& PGI-10), Forschungszentrum J\"ulich GmbH, 52428 J\"ulich, Germany}
\affiliation{JARA-Fundamentals of Future Information Technology, Forschungszentrum J\"ulich and RWTH Aachen University, 52062 Aachen, Germany}

\author{Amirehsan Boreiri}
\affiliation{Institute for Research in Electronics and Applied Physics and Joint Quantum Institute, University of Maryland, College Park, Maryland 20742, USA}
\affiliation{Department of Electrical and Computer Engineering, University of Maryland, College Park, MD 20742, USA}

\author{Douglas L. Irving}
\affiliation{Department of Materials Science and Engineering, North Carolina State University, Raleigh, North Carolina 27695, USA}

\author{Alexander Pawlis}
\affiliation{Peter-Gr\"unberg Institute (PGI-9 \& PGI-10), Forschungszentrum J\"ulich GmbH, 52428 J\"ulich, Germany}
\affiliation{JARA-Fundamentals of Future Information Technology, Forschungszentrum J\"ulich and RWTH Aachen University, 52062 Aachen, Germany}

\author{Edo Waks}
\email{edowaks@umd.edu}
\affiliation{Institute for Research in Electronics and Applied Physics and Joint Quantum Institute, University of Maryland, College Park, Maryland 20742, USA}
\affiliation{Department of Electrical and Computer Engineering, University of Maryland, College Park, MD 20742, USA}


\begin{abstract}
Semiconductor hole spins offer a pathway to extended coherence times by decoupling from nuclear magnetic noise, while their spin-orbit coupling enables fast all-electrical control. In ZnSe, however, realizing this potential has been limited by p-doping challenges. Here, we circumvent this limit by optically activating acceptors within the ZnSe quantum well. We isolate a single-hole spin bound to a shallow acceptor, confirmed by antibunching and accessed via the fast (244 ps) radiative recombination of a bound exciton.  Magnetic and Raman spectroscopy of the ground state reveal an effective hole g-factor of 0.7 and an optical resonance linewidth of 26.7 GHz. Complementary first-principles simulations, together with the experimental results, provide evidence that points toward nitrogen as the most likely acceptor impurity. These results introduce a promising new platform for optically active spin qubits and single-photon sources in ZnSe.

\end{abstract}

\maketitle

\input{sections/Section0.tex}  

\input{sections/SectionI.tex}  

\input{sections/SectionII.tex}

\input{sections/SectionIII.tex}

\input{sections/Conclusion.tex}

\input{sections/acknowledgements.tex}

\bibliographystyle{ieeetr}   
\bibliography{refs.bib}          


\end{document}


\title{Supplemental Material for "Optically active single hole spin in ZnSe"}

\author{Amirehsan Alizadehherfati}
\email{herfati@umd.edu}
\affiliation{Institute for Research in Electronics and Applied Physics and Joint Quantum Institute, University of Maryland, College Park, Maryland 20742, USA}
\affiliation{Department of Electrical and Computer Engineering, University of Maryland, College Park, MD 20742, USA}

\author{Yuxi Jiang}
\affiliation{Institute for Research in Electronics and Applied Physics and Joint Quantum Institute, University of Maryland, College Park, Maryland 20742, USA}
\affiliation{Department of Electrical and Computer Engineering, University of Maryland, College Park, MD 20742, USA}

\author{Kelsey J. Mirrielees}
\affiliation{Department of Materials Science and Engineering, North Carolina State University, Raleigh, North Carolina 27695, USA}

\author{Nils von den Driesch}
\affiliation{Peter-Gr\"unberg Institute (PGI-9 \& PGI-10), Forschungszentrum J\"ulich GmbH, 52428 J\"ulich, Germany}
\affiliation{JARA-Fundamentals of Future Information Technology, Forschungszentrum J\"ulich and RWTH Aachen University, 52062 Aachen, Germany}

\author{Christine Falter}
\affiliation{Peter-Gr\"unberg Institute (PGI-9 \& PGI-10), Forschungszentrum J\"ulich GmbH, 52428 J\"ulich, Germany}
\affiliation{JARA-Fundamentals of Future Information Technology, Forschungszentrum J\"ulich and RWTH Aachen University, 52062 Aachen, Germany}

\author{Yurii Kutovyi}
\affiliation{Peter-Gr\"unberg Institute (PGI-9 \& PGI-10), Forschungszentrum J\"ulich GmbH, 52428 J\"ulich, Germany}
\affiliation{JARA-Fundamentals of Future Information Technology, Forschungszentrum J\"ulich and RWTH Aachen University, 52062 Aachen, Germany}

\author{Amirehsan Boreiri}
\affiliation{Institute for Research in Electronics and Applied Physics and Joint Quantum Institute, University of Maryland, College Park, Maryland 20742, USA}
\affiliation{Department of Electrical and Computer Engineering, University of Maryland, College Park, MD 20742, USA}

\author{Douglas L. Irving}
\affiliation{Department of Materials Science and Engineering, North Carolina State University, Raleigh, North Carolina 27695, USA}

\author{Alexander Pawlis}
\affiliation{Peter-Gr\"unberg Institute (PGI-9 \& PGI-10), Forschungszentrum J\"ulich GmbH, 52428 J\"ulich, Germany}
\affiliation{JARA-Fundamentals of Future Information Technology, Forschungszentrum J\"ulich and RWTH Aachen University, 52062 Aachen, Germany}

\author{Edo Waks}
\email{edowaks@umd.edu}
\affiliation{Institute for Research in Electronics and Applied Physics and Joint Quantum Institute, University of Maryland, College Park, Maryland 20742, USA}
\affiliation{Department of Electrical and Computer Engineering, University of Maryland, College Park, MD 20742, USA}


\maketitle
\onecolumngrid

\input{sectionsSI/Methods.tex}  

\input{sectionsSI/Sectionfirstprinciple.tex}

\input{sectionsSI/SectionII.tex}

\input{sectionsSI/SectionI.tex}

\input{sectionsSI/Section0.tex}

\bibliographystyle{ieeetr}   
\bibliography{refsSI.bib}          


%% file: preamble.tex
\usepackage{amsthm}
\usepackage{mathtools}
\usepackage{physics}
\usepackage{xcolor}
\usepackage{graphicx}
\usepackage[left=23mm,right=13mm,top=35mm,columnsep=15pt]{geometry} 
\usepackage{adjustbox}
\usepackage{placeins}
\usepackage[T1]{fontenc}
\usepackage{lipsum}
\usepackage{csquotes}

%% file: sections/Section0.tex
\section{Introduction}

Impurity-bound excitons in ZnSe provide bright single-photon emission and a promising spin-photon platform for quantum photonics \cite{Greilich2012,Heisterkamp2015}. The ZnSe host has a direct wide bandgap ($\approx$ 2.8 eV), offering a high Debye-Waller factor ($\approx$ 94$\%$) \cite{Jiang2024}  that yields efficient radiative transitions \cite{Steiner1985}. Moreover, ZnSe has a nearly nuclear-spin-free environment, which can be isotopically purified to further extend the spin coherence \cite{Kopteva2019,Pawlis2019MBE,Kirstein2021Extended}. Donor-bound excitons, particularly from Cl and F, have been isolated at the single impurity level \cite{Kutovyi2022Efficient,De-Greve2010-fi}. Experiments have demonstrated indistinguishable photons from independent donors \cite{Sanaka2009Indistinguishable,Sanaka2012Entangling}, integration with nanophotonic structures to enhance light matter interaction \cite{doi:10.1021/acsphotonics.3c01540,Qiao2024}, ground-state electron initialization \cite{Kim2014-od} and electrical control of donor spins \cite{alizadehherfati2025electrical}. 

 In comparison, acceptor-bound excitons in ZnSe remain scarcely explored in the single-impurity limit \cite{Tews1979,Zhang1994}, despite offering distinct advantages for quantum information \cite{delteil2016generation,hogg2025fast}. Hole spins benefit from a p-orbital symmetry that suppresses hyperfine interactions with nuclear spins, promising extended coherence times \cite{PhysRevLett.106.027402,prechtel2016decoupling}. Additionally, their strong spin-orbit coupling enables fast all-electrical spin manipulation \cite{wang2022ultrafast}, a key requirement for scalable quantum architectures. Furthermore, the stronger localization of the bound hole in acceptors inside the quantum well, compared to well-known quantum dot systems, results in a more reproducible potential landscape and higher homogeneity \cite{van2018readout}. Although anti-bunching from nitrogen acceptors in p-doped ZnSe has been demonstrated \cite{strauf2002quantum}, progress has been limited, in part, by the difficulty of achieving reliable p-doping in ZnSe quantum wells \cite{PhysRevLett.74.1131,PhysRevB.45.10965,DePuydt1989,Cheong1995,Haase1990}.

In this letter, we isolate a single hole spin bound to a nitrogen acceptor within ZnSe quantum well. We achieve this by neutralizing the optically dark acceptor via above-bandgap optical charge injection \cite{Dusanowski2022}. The necessity of this injection mechanism is confirmed through quasi-resonant excitation experiments, where the acceptor emission is present only with the co-application of an above-band pump. Based on direct probing of the spin dynamics using resonant Raman spectroscopy, we observe efficient spin-flip process as an indication of spin initialization.  These findings establish hole spin in ZnSe as an interesting study candidate for quantum nanophotonic and spintronic applications. 

\begin{figure}[h]
  \centering
\includegraphics[width=\columnwidth]{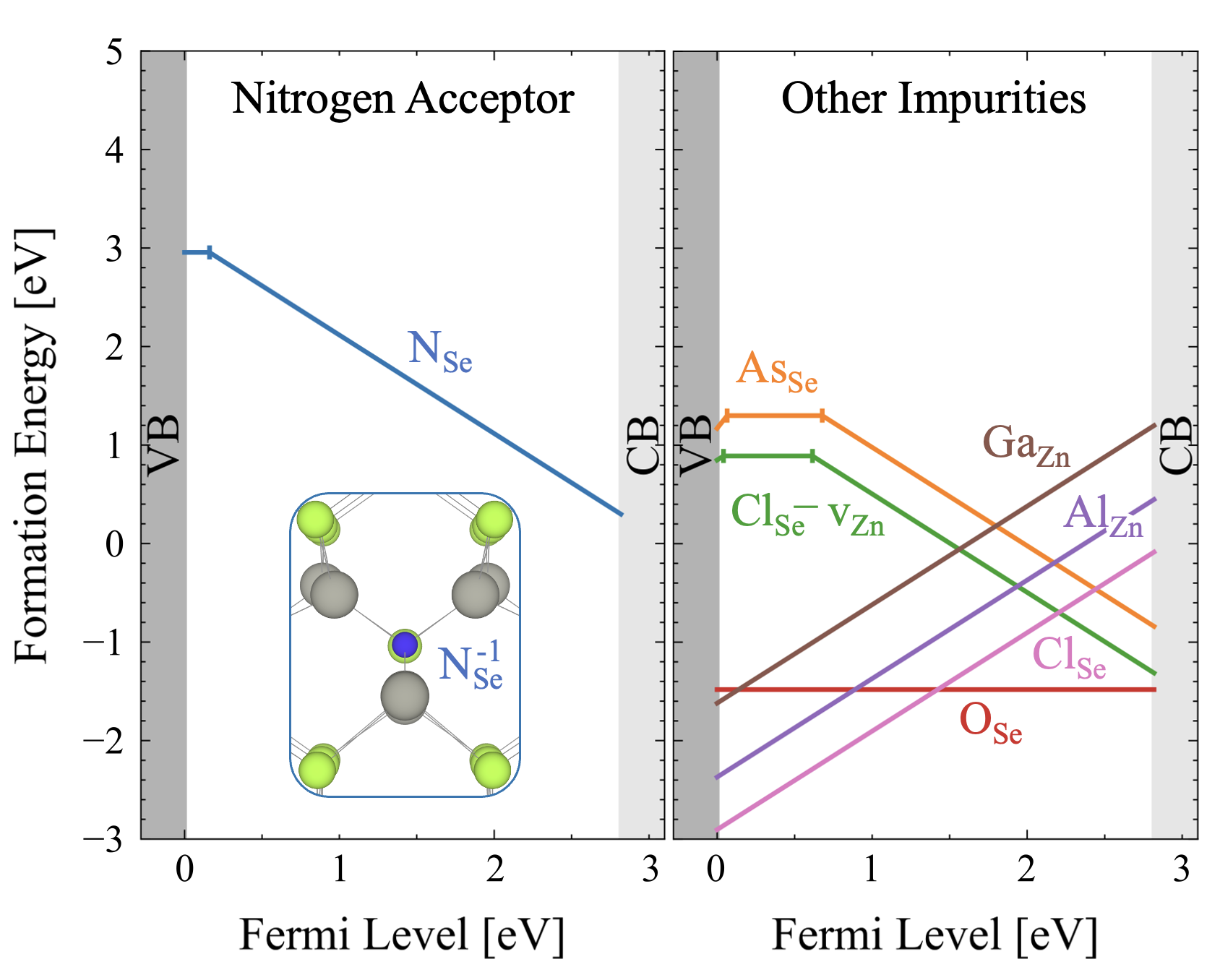}
  \caption{\textbf{Formation energy diagrams.} Left panel: calculated formation energies of the nitrogen substitutional in ZnSe. Right panel: calculated formation
energies of other impurities in ZnSe.
}
\label{fig:Figfirst}
\end{figure}

\begin{figure*}[t]
  \centering
\includegraphics[width= 2\columnwidth]{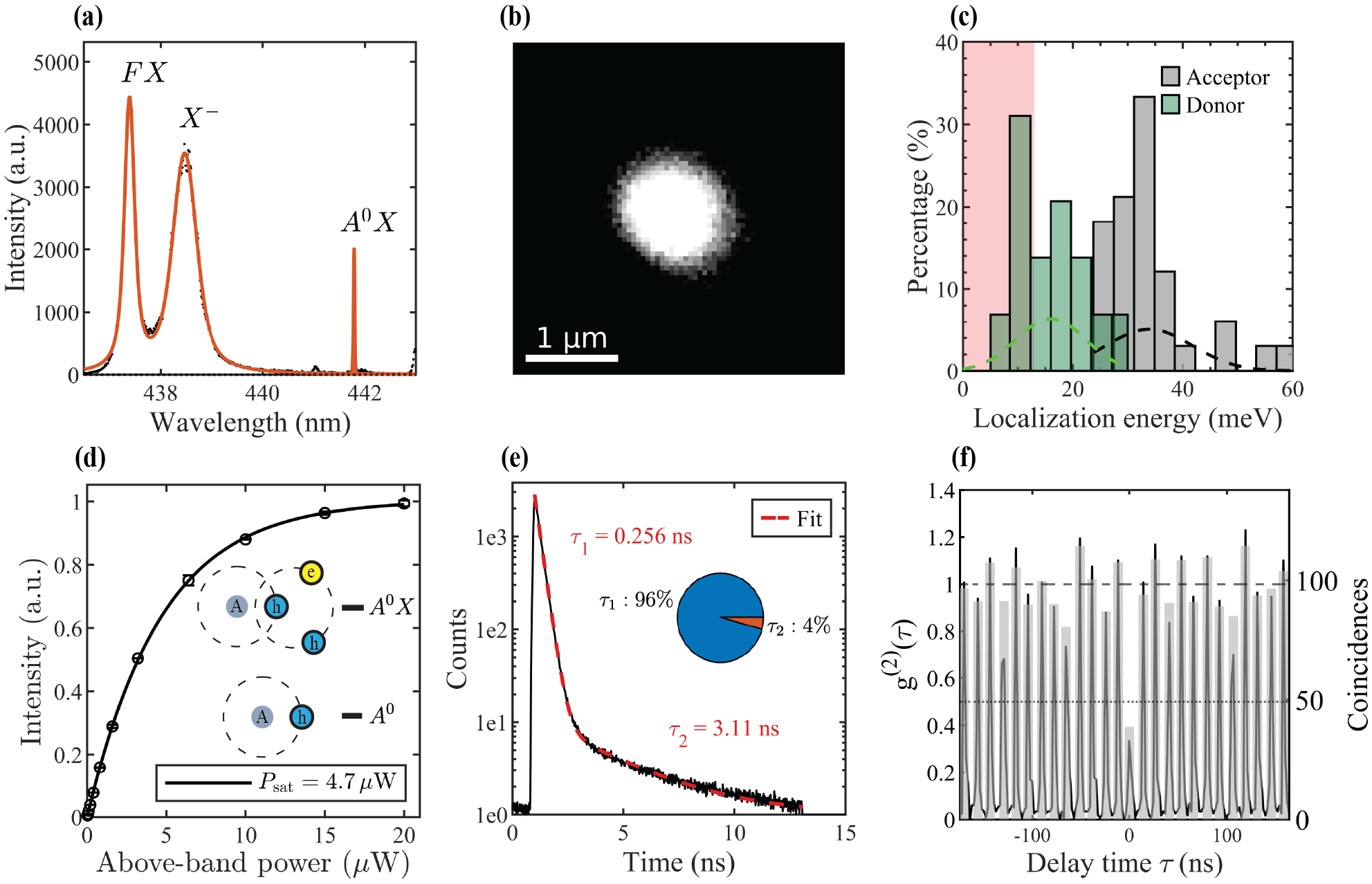}
  \caption{\textbf{Photoluminescence characterization.} 
  (a) Sample spectrum containing three main components, free exciton (labeled as FX), negatively charged trion (labeled as $X^-$) and bound exciton line (labeled as $A^0X$) (b) Photoluminescence spatial image of the bound exciton line, achieved by selective filtering the sharp peak. (c) Histogram of bound excitons localization energies, indicating two distributions of peaks for acceptors (gray) and donors (green). We collected the data from 30 different acceptor sites and verified them through magnetospectroscopy. The shaded region indicates the region masked by the strong trion envelope in our sample. (d)  Power-dependent photoluminescence excitation experiment showing two-level saturating behavior, with embedded level structure of acceptor bound exciton state $A^0X$ and neutral acceptor $A^0$. (e) Pulsed excitation measurement shows double exponential decay with a fast component of 256 ps, related to the excited state lifetime and a slow contribution of 3.11 ns. (f) Time correlation measurement between two detectors, obtained from the acceptor line, showing antibunching and single photon emission.  
}
\label{fig:Fig0}
\end{figure*}

%% file: sections/SectionI.tex
\section{Photoluminescence Characterization}

The device under study is a ZnMgSe/ZnSe/ZnMgSe quantum well grown on a GaAs substrate (Methods A), featuring a central chlorine $\delta$-doped layer \cite{Ohkawa1987}. While the chlorine atoms predominantly act as substitutional shallow donors ($\mathrm{Cl}_{\mathrm{Se}}^{+1}$) \cite{dosSantos2011,Poykko1998}, the quantum well also hosts a natural minority acceptor population. Crucially, the distinct binding energies of these species provide sufficient spectral separation to investigate both within the same sample. However, the identity and nature of these minority acceptors remain poorly understood, motivating a systematic investigation of which impurities can form under typical growth conditions.

To identify the source of the acceptor population, we perform first-principles calculations for a wide breadth of impurities identified as likely being present in the growth reactor used to synthesize ZnSe. The impurities considered include nitrogen, arsenic, oxygen, chlorine \cite{wu2022defect}, aluminum \cite{mirrielees2025assessing}, and gallium \cite{mirrielees2025assessing} as well as native defects \cite{Wu2022NativeZnSe}. Formation energies of these impurities are calculated using hybrid functional density functional theory simulations. Simulation details can be found in the supplementary material section I. Among the impurities considered in this work, nitrogen substituting on the Se site (N${_\mathrm{Se}}$), arsenic substituting on the Se site (As${_\mathrm{Se}}$), and the chlorine–vacancy complex (Cl${_\mathrm{Se}}$–v${_\mathrm{Zn}}$) emerge as the most likely acceptor-type defects in ZnSe. Their calculated formation energies are shown in Fig. \ref{fig:Figfirst}, with the corresponding thermodynamic transition levels pronounced with tick marks in the formation energy diagram. Among the defects considered, N$_\mathrm{Se}$ forms the shallowest acceptor state in ZnSe with an ionization energy of 0.16 $\mathrm{eV}$, whereas the other candidates introduce substantially deeper levels in the bandgap (0.68 $\mathrm{eV}$ for As${_\mathrm{Se}}$ and 0.62 $\mathrm{eV}$ for Cl${_\mathrm{Se}}$–v${_\mathrm{Zn}}$). Meanwhile, Ga$_\mathrm{Zn}$, Al$_\mathrm{Zn}$, and Cl$_\mathrm{Se}$ are shallow donors and O$_\mathrm{Se}$ is neutral across the bandgap. Due to the dominant chlorine donor population and intrinsic n-type character of ZnSe, the Fermi level is expected to lie above the acceptor level, leaving the acceptors in their ionized charge state (-1). Having identified nitrogen as the shallowest impurity acceptor, we now turn to low temperature optical measurements to search for experimental evidence of its presence in the sample. 



We load the sample into a closed-loop cryostat with a base temperature of 3.6 K and utilize a home-built confocal microscope (Methods B), similar to Ref. \cite{alizadehherfati2025electrical}.  We perform photoluminescence spectroscopy under above-band excitation at \SI{405}{\nano m}. Fig. \ref{fig:Fig0}(a) shows a sample spectrum,  containing the free exciton (labeled as FX), negatively charged trion (labeled as $X^-$), and a sharp bound exciton line (labeled as $A^0X$). We define the localization energy as the energy difference between the free and bound exciton emission, which yields a value of 28.3 meV for this line.  Fig. \ref{fig:Fig0}(b) shows an image of the photoluminescence spatial map from this line, recorded by filtering the sample emission through a narrow band pass filter centered at 441.6 nm with a pass band of about 1.7 nm, demonstrating the localized nature of this emission. We observe similar emission peaks across the sample, with localization energies exceeding 25 meV. The distribution of these peaks is plotted in Fig. \ref{fig:Fig0}(c), alongside reproduced data for Cl donors from Ref. \cite{Karasahin2022} for comparison. The distinct range of observed localization energies indicates that the emission origin differs from that of typical chlorine-bound excitons. 
We attribute these emissions to the radiative recombination of excitons bound to shallow acceptors \cite{Merz1973,Hu1993,Chen1995}. Returning to the candidates identified by our defect calculations, the average localization energy of $33.9\pm 7.9$ meV is most consistent with N${_\mathrm{Se}}$, whose shallow acceptor level aligns well with the observed values, while other candidates introduce deeper levels expected to yield substantially larger localization energies. Based on this, nitrogen is the most likely candidate (see supplemental material section II) \cite{Dean1983,PhysRevB.48.17827}. 

To verify that the emission originates from an isolated two-level system, we perform power-dependent photoluminescence measurements. Fig.~\ref{fig:Fig0}(d) displays the integrated intensity of the bound exciton emission as a function of above-band excitation power, fitted with a standard two-level saturation model (solid line). The excellent agreement between the model and experimental data confirms that the emission arises from the recombination of an exciton bound to a neutral acceptor, forming an effective two-level system. The relevant energy level scheme, depicting the excited ($A^0X$) and ground ($A^0$) states, is illustrated in the inset. The fit yields a saturation power of \SI{4.7}{\micro W}, consistent with typical values for single-bound excitons in unstructured ZnSe quantum wells \cite{Karasahin2022}. 

  To characterize the lifetime of the bound exciton excited state, we perform time-resolved fluorescence measurements using 4 ps above-band laser pulses. Fig.~\ref{fig:Fig0}(e) displays the histogram of photon arrival times fitted with a biexponential decay model. The extracted dynamics reveal a dominant fast component with an average decay time of \SI{256}{\pico s} (see supplemental material section III) and a minor slow component of \SI{3.11}{\nano s}. We attribute the fast decay to the radiative lifetime of the $A^0X$ excited state, consistent with values reported for nitrogen acceptors \cite{strauf2002quantum,kothandaraman1997temperature}. The slow decay tail, corresponding to $4\%$ of total decay, is ascribed to repopulation dynamics caused by nearby trap states or dark states feeding the acceptor transition \cite{Karasahin2022}. 

We verify the single-photon nature of the acceptor-bound exciton emission via pulsed laser autocorrelation measurements. Fig.~\ref{fig:Fig0}(f) displays the second-order correlation function, $g^{(2)}(\tau)$, derived from the photon arrival times. The plot shows both the raw coincidence histogram (black lines) and the integrated peak areas (gray bars) for each pulse interval. We observe clear antibunching at zero delay, yielding a raw value of $g^{(2)}(0) = 0.39 \pm 0.09$. To account for uncorrelated background noise, primarily arising from the trion envelope tail enhanced by the above-band excitation, we apply a background correction based on the measured signal-to-total intensity ratio of $80\%$ \cite{doi:10.1021/acsphotonics.3c01540}. This yields a corrected value of $g^{(2)}(0) \approx 0.05$, confirming that the emission originates from a single isolated defect. These results align with previous observations of antibunching from nitrogen acceptors \cite{strauf2002quantum}.

%% file: sections/SectionII.tex
\section{Photoluminescence Excitation Measurement}

\begin{figure}[h]
  \centering
\includegraphics[width=\columnwidth]{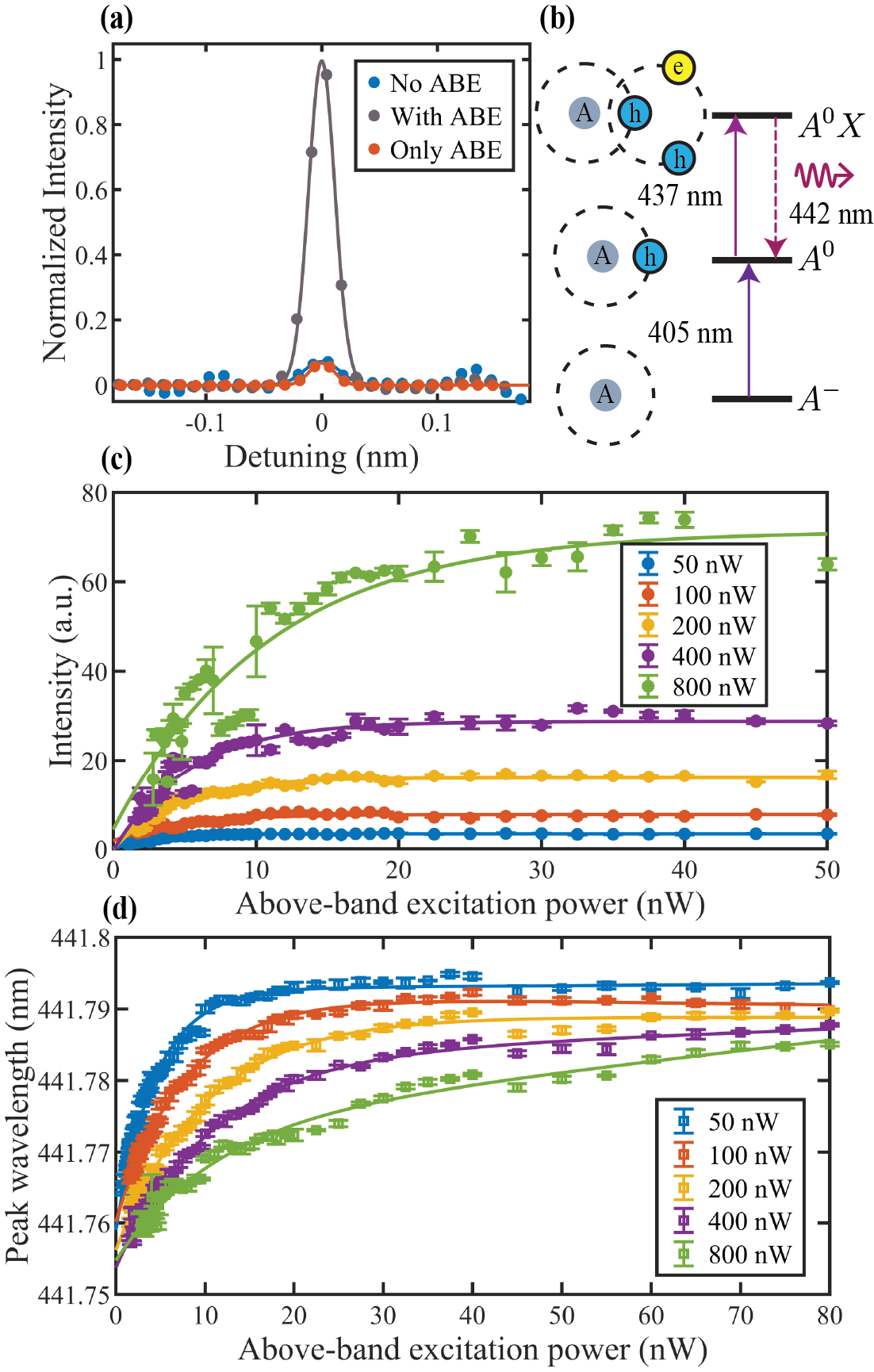}
  \caption{\textbf{Photoluminescence excitation measurement of a single acceptor-bound exciton.} (a) Spectrum of emission in the absence and presence of above-band excitation (labeled as ABE). (b) Level structure and charge states of the acceptor system. The arrows show the simplified dominant contribution of each laser. (c) Intensity of emission as a function of above-band excitation for different powers of resonant excitation (see legend).  (d) Central wavelength of emission as a function of above-band power for different powers of resonant excitation. The trend shows a saturating shift toward higher wavelengths.
}
\label{fig:Fig1}
\end{figure}

We perform photoluminescence excitation spectroscopy on the acceptor-bound exciton transition, similar to the method in Ref. \cite{alizadehherfati2025electrical}. Unlike in donor studies, we detect negligible signal unless a weak above-band pump is simultaneously applied. Fig.~\ref{fig:Fig1}(a) compares the photoluminescence excitation spectra of the acceptor emission with and without this additional illumination at a fixed resonant power of \SI{200}{\nano W}. The weak above-band power (\SI{20}{\nano W}) contributes negligible direct emission. 

We attribute this observation to the optical manipulation of the acceptor charge state via the generation and capture of free holes. This process is shown in Fig.~\ref{fig:Fig1}(b). At equilibrium, the Fermi level is pinned above the acceptor energy, maintaining the impurities in the optically dark ionized state ($A^-$). The weak above-band illumination generates excess holes, which can be efficiently captured by individual ionized acceptors. Once neutralized ($A^0$), the acceptors can capture resonantly created free excitons via the \SI{437}{\nano m} laser to form the acceptor-bound exciton complex ($A^0X$), thereby activating the emission.
\begin{figure*}[t]
  \centering
\includegraphics[width= 430.4115  pt]{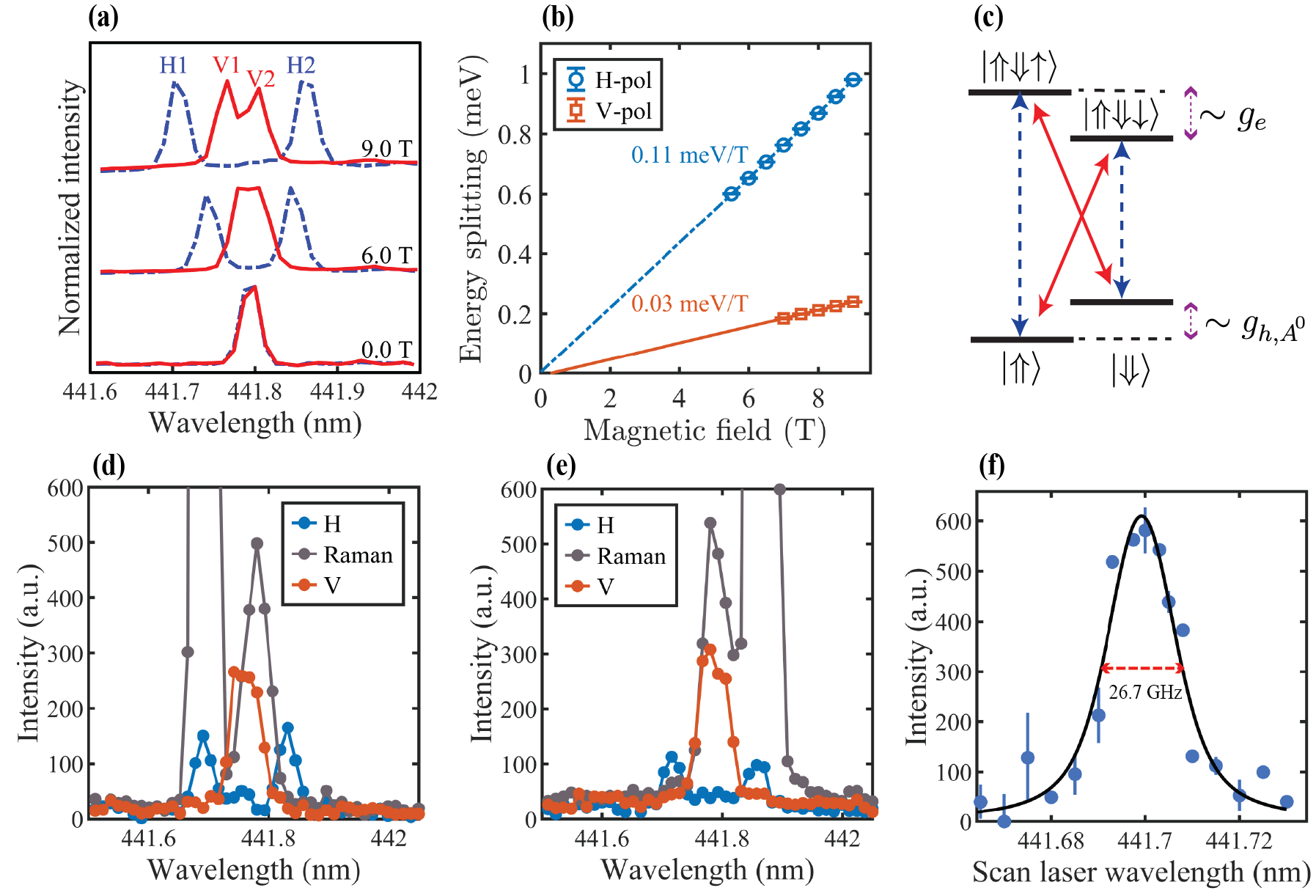}
  \caption{\textbf{Magnetic and Raman spectroscopy.} (a) Photoluminescence spectrum as a function of magnetic field. (b) Energy splitting of the spectral peaks as a function of magnetic field. The blue and red data points correspond to horizontal (H) and vertical (V) polarizations, respectively. Solid lines represent linear fits extracting the Zeeman splitting rate for each polarization state. (c) Level structure of the acceptor ground and excited state based on extracted g-values. (d-e) Raman sideband (cross-polarized) appears at the partner transition energy when pumping the opposite polarization. For this measurement, we fix the external magnetic field to \SI{6}{T}. (f) Integrated intensity of Raman sideband as a function of tunable laser wavelength, showing linewidth of \SI{111}{\micro eV}.
} 
\label{fig:Fig3}
\end{figure*}

To investigate the charge dynamics, we analyze the bound exciton intensity as a function of above-band power for varying resonant excitation levels, shown in Fig. \ref{fig:Fig1}(c). The solid lines are fits to the experimental data using a saturation model. We find that increasing the resonant drive not only enhances the saturated emission intensity but also shifts the saturation threshold to higher powers. This dual dependence is fully described by the phenomenological rate-equation model derived in the supplemental material section IV. The analysis shows that the resonant laser does not simply drive the optical transition but simultaneously induces an active ionization process. This creates a dynamical competition where the above-band pump works to neutralize the acceptor, while the resonant laser counteracts this by driving an ionization process. We attribute this to the Auger-assisted autoionization process, where the resonant laser ejects the hole and resets the acceptor to its ionized state \cite{PhysRev.176.993}.

Beyond the emission intensity, the central wavelength serves as a sensitive probe of the local electrostatic environment. Figure \ref{fig:Fig1}(d) displays the evolution of the emission wavelength as a function of above-band power for varying resonant excitation levels. The solid lines represent numerical fits to the experimental data using a saturation model. We observe a distinct redshift that saturates with increasing above-band power, where the saturation threshold shifts to higher values as the resonant drive is increased. We attribute this redshift to a DC Stark effect induced by the changing electric field landscape of nearby charge traps. As the above-band pump modifies the occupancy of these traps, the local electric field shifts, tuning the central wavelength of acceptor-bound exciton transition \cite{houel2012probing}. The dependence of the saturation threshold on resonant power confirms that the resonant laser also actively modifies this trap occupancy, further validating the competitive charge dynamics model discussed earlier. 

%% file: sections/SectionIII.tex
\section{MAGNETOSPECTROSCOPY Characterization and Raman Spectroscopy}
\label{sec:develop}

To investigate the spin properties of the bound exciton complex, we perform polarization-resolved magnetospectroscopy. Fig. \ref{fig:Fig3}(a) shows the photoluminescence signal for fields up to 9T in the Voigt configuration. The magnetic field lifts the degeneracy of the electron and hole in both ground and excited states \cite{astakhov2002binding}, producing a linear Zeeman splitting in each polarization, labeled as H and V. 
Fig. \ref{fig:Fig3}(b) shows the energy splitting between co-polarized transitions as a function of magnetic field for each polarization. We extract each energy by fitting a double Voigt profile to the data and plot these separations as a function of magnetic field.

To extract the electron and hole g-factors, we analyze the Zeeman splitting based on the energy level structure shown in Fig. \ref{fig:Fig3}(c). The ground state ($A^0$) consists of a single hole, and its levels (labeled as $\ket{\Uparrow}$ and $\ket{\Downarrow}$) split according to the hole g-factor, $g_h$. The excited state is the acceptor-bound exciton ($A^0X$), a three-particle complex. In this state, the two holes are assumed to form a spin singlet, leaving the total spin of the complex determined by the single electron. Thus, its levels (labeled as $\ket{\Uparrow\Downarrow\uparrow}$
and $\ket{\Uparrow\Downarrow\downarrow}$) split with the electron g-factor, $g_e$. From the polarization-resolved Zeeman splitting, we obtain $g_e = 1.17\pm 0.01$ for excited state electron and $g_h=0.70\pm 0.01$ for ground state hole. These two values are in close agreement with previous studies on nitrogen acceptor-bound excitons in p-doped ZnSe \cite{Stadler1995,Hoffmann1996}.

To verify the interconnected energy level structure of the bound exciton complex, we perform spin-flip Raman spectroscopy by pumping each Zeeman-split transition with \SI{100}{\nano W} and observing the emission at the cross-polarized partner transition \cite{PhysRevB.56.6889,PhysRevB.55.1607}. Notably, simultaneous application of above-band pump (\SI{100}{\nano W}) is essential for this measurement. In the absence of this above-band illumination, no emission signal is detected. Crucially, the required above-band power is an order of magnitude higher than the saturation threshold for charge neutralization determined via photoluminescence excitation spectroscopy (\SI{10}{\nano W}). This disparity indicates that the additional optical flux is necessary to actively randomize the spin population. The rapid randomization counteracts the optical pumping induced by the strong resonant drive, preventing the system from becoming trapped in a dark spin state \cite{Kim2014-od}. Figs. \ref{fig:Fig3}(d) and \ref{fig:Fig3}(e) display the results, where the blue (H) and red (V) traces show the reference photoluminescence spectra. The gray trace (labeled as Raman) shows the cross-polarized (V) collected emission while the pump laser is tuned on resonance with an H-polarized transition. The appearance of an enhanced emission sideband, precisely at the energy of the corresponding V-polarized transition, is the signature of an efficient spin-flip process. This observation confirms that the transitions are interconnected through a $\Lambda$-system and provides qualitative evidence for optical hole spin initialization. 

To determine the optical resonance linewidth, we scan the tunable laser through an H-polarized transition and record the intensity of the V-polarized sideband for a fixed resonant and above-band excitation.  Fig. \ref{fig:Fig3}(f) shows the total intensity of the sideband as a function of resonant laser wavelength. The solid line shows a Voigt fit to the experimental data. From the fit, we extract a linewidth of 111 $\mu$eV (corresponding to 26.7 GHz), which is about 40 times larger than the lifetime-limited linewidth (0.65 GHz). We attribute this to spectral diffusion caused by charge fluctuations in the local trap environment, a phenomenon also observed in donor-bound excitons \cite{alizadehherfati2025electrical,jiang2024generation}. Charge injection and stabilization via electrical gating offer a promising pathway to mitigate this decoherence and narrow the linewidths.

%% file: sections/Conclusion.tex
\section{Conclusion}
\label{sec:develop}

In summary, we report optical activation and isolation of shallow acceptor-bound hole spins in ZnSe quantum well. By employing a dual-laser excitation scheme, we overcome the intrinsic donor compensation to neutralize the ionized acceptors. Magnetic and Raman spectroscopy verify the formation of a $\Lambda$-system suitable for spin manipulation. However, the Raman optical linewidth ($\sim 26.7$ GHz) remains significantly broader than the transform limit, indicating that the coherence is currently limited by spectral diffusion caused dominantly by a fluctuating charge environment. 
Future efforts should focus on stabilizing this environment and implementing coherent control. A practical next step is the integration of the quantum well into a Schottky diode structure on a p-doped GaAs substrate \cite{kuhlmann2013charge}. This architecture would allow for the application of vertical electric fields to suppress charge noise via carrier depletion while electrically neutralizing the acceptor. Furthermore, such electrical gating could be leveraged to drive fast electric dipole spin resonance, exploiting the strong spin-orbit coupling of the holes to achieve coherent spin manipulation without the need for local magnetic fields \cite{watzinger2018germanium}. To enable high-fidelity coherence measurements, photon collection efficiency can be enhanced by coupling the sample to a solid immersion lens. This improved signal-to-noise ratio would facilitate the observation of coherent population trapping \cite{PhysRevB.103.115412}, providing a direct measurement of the spin coherence time. Finally, coupling these emitters to optical microcavities offers a pathway to exploit the Purcell effect for cavity-QED applications, effectively increasing the radiative rate and improving photon indistinguishability \cite{tomm2021bright}.

%% file: sections/acknowledgements.tex
\section*{Acknowledgments}
The Waks group would like to acknowledge support from the AFOSR (grant \#FA95502010250, grant \#FA95502410266, and grant \#FA95502310667). The Irving group acknowledges support for this work from AFOSR Grant FA9550-21-1-0383. The Pawlis group would
like to acknowledge support from the Deutsche Forschungsgemeinschaft (DFG, German Research
Foundation) under Germany's Excellence Strategy - Cluster of Excellence Matter and Light for Quantum Computing  (ML4Q) EXC 2004/1-390534769. The funder played no role in study design, data collection, analysis and interpretation of data, or the writing of this manuscript.

\section*{DATA AND MATERIALS AVAILABILITY}
All of the data that support the findings of this study are reported in the main text and Supplementary Materials. Source data are available from the corresponding authors on reasonable request.

\section*{Competing interests}
The authors declare that they have no competing interests.

\section*{Author contributions} A.A., A.P., and E.W. conceived the experiment. C.F., N.D., and Y.K. fabricated the device. A.A.
performed the experiment. Y.J. and A.B. supported setting up the experiment. A.A. analyzed the experimental data. K.M. and D.I. provided the first principle simulations and analysis of this theoretical data. A.A., K.M. and E.W. prepared the manuscript. All authors discussed the results and confirmed the manuscript. E.W. and A.P. supervised the experiment.

%% file: sectionsSI/Methods.tex
\section*{Methods}\label{Methoods}
\subsection{Device description}\label{subsec:device}

The ZnMgSe/ZnSe/ZnMgSe quantum well heterostructure was grown by molecular-beam epitaxy on GaAs. The ZnSe QW is \(4.8\,\mathrm{nm}\) thick and is sandwiched between \(27.7\,\mathrm{nm}\) ZnMgSe barriers above a \(14\,\mathrm{nm}\) ZnSe buffer layer on GaAs.  Chlorine donors were incorporated into the QW by \(\delta\)-doping at a sheet density of \(\sim 10^{10}\,\mathrm{cm}^{-2}\). A 40 nm thick oxide capping layer was grown, acting as a protective layer for ZnSe well.

\subsection{Experimental Setup}\label{subsec:setup}
The sample is mounted in a closed loop cryostat (Bluefors) working at \SI{3.6}{K} equipped with a superconducting magnet providing fields up to 9 T in Voigt configuration (American Magnetics). A free-space confocal microscope is used for excitation and collection, with a \(0.68\)-NA objective and a focal length of 3 mm. Above-band excitation at \(405\,\mathrm{nm}\) is provided by a diode laser (Thorlabs, LP405\textendash SF10). The resonant excitation and the laser scanning are performed using a tunable diode laser (TOPTICA, DL pro), a frequency-doubled pulsed
Ti:sapphire laser (Mira) and a scanning Fabry-Perot interferometer (Thorlabs, SA30-47). We used a \SI{600}{MHz} resolution wavemeter (HighFinesse, Angstrom WS/6) to monitor the central wavelength of resonant laser. The resonant beam is delivered through polarization-maintaining single-mode fiber (PM-SMF) and then passes a linear polarizer, half-wave plate, and quarter-wave plate to prepare the incident polarization for cross-polarized detection. The collected signal passes through a different set of quarter-wave plate, half-wave plate, and linear polarizer, and is then collected by a polarization-maintained single mode fiber. Spectra are acquired with a Princeton Instruments spectrometer comprising a monochromator (1800\,g/mm grating) and a 1340 pixel CCD camera. For correlation and time-resolved measurements, the collected signal is split by a \(50{:}50\) fiber beam splitter and detected on two fiber-coupled superconducting single-photon detectors (QuantumOpus). Detector outputs are time-tagged by a HydraHarp400 (PicoQuant). We observe a response time (jitter) of about 
\SI{30}{\pico s} using 4 ps pulses.

%% file: sectionsSI/Sectionfirstprinciple.tex
\section{First principle Results}

Hybrid exchange-correlation density functional theory (DFT) with the functional of Heyd, Scuseria, and Ernzerhof (HSE)\cite{heyd2003hybrid,heyd2004efficient,heyd2005energy} was used explore possible acceptor defects in zinc-blende ZnSe. 
An exact-exchange fraction of 0.3445 for ZnSe simulations resulted in a bandgap of 2.82 eV.\cite{wu_native_2022}
Pseudopotentials were treated within the projector augmented wave approach, and Zn and Se pseudopotentials contained 12 and 6 valence electrons.\cite{blochl1994projector,kresse1999ultrasoft}
For calculating formation energy diagrams (FEDs), defect supercells having a single substitutional defect such as N substituting the Se site (N$_{\rm Se}$) contained a total of 72 atoms.
Larger supercells of up to 96 atoms were used for extended point defects, such as in the case of the chlorine--vacancy complex (Cl$_{\rm Se}$-- v$_{\rm Zn}$).
Atoms within a 5 Å radius around the defect were allowed to relax while atoms near supercell edges were frozen.
A Monkhorst-Pack 2x2x2 k-point mesh was used for calculating energies of defect and bulk supercells.

Formation energies for all point defects in this work were calculated in the following way:
\begin{equation}
    E_{D^q}^f=E_{D^q}^{\rm tot}-E_{\rm bulk}^{\rm tot}-\sum_in_i\mu_i+q(\mu_e+E_v)+E_{D^q}^{\rm corr}
\end{equation}
where $E_{D^q}^f$ is the defect formation energy, $E_{D^q}^{\rm tot}$ and $E_{\rm bulk}^{\rm tot}$ are the total DFT energies of the supercell containing the defect and the pristine bulk, respectively.
The chemical potential, $\mu_i$, has been calculated for each elemental species, $i$,  with $n_i$ being the number of atoms of $i$  added or removed from the pristine bulk cell.
Defects were calculated at multiple charge states, $q$, $\mu_e$ is the Fermi level relative to the valence band maximum $E_v$, and $E_{D^q}^{\rm corr}$ is the finite size correction based on that of Kumagai and Oba.\cite{kumagai2014electrostatics}
Point defects are managed and analyzed via an informatics and software suite.\cite{Baker2019}

%% file: sectionsSI/SectionII.tex
\section{Acceptor Localization energy in QW}
To validate the assignment of this peak to nitrogen based on localization energy, we calculated the expected value by combining bulk binding trends with a confinement scaling factor derived from donors. We first determine the bulk baseline using the extracted ionization energy of nitrogen acceptors in ZnSe ($E_A \approx 160$ meV) from our first-principles study. Applying the established Haynes' rule for neutral acceptors in II-VI compounds ($E_{loc}/E_A \approx 0.1$) \cite{halsted1965photoluminescence,PhysRevLett.4.361} yields a predicted bulk localization energy of $E_{loc}^{bulk} \approx 16$ meV \cite{dean1983ionization}. To account for the quantum well environment, we determine a confinement enhancement factor based on variational models, resulting in an enhancement of 1.9 for a Bohr radius of 3 nm in a 4.8 nm quantum well \cite{bastard1981hydrogenic}. This is consistent with the chlorine donors as a reference, exhibiting an enhancement of around 2 from bulk ZnSe ($\sim 7$ meV \cite{PhysRevB.23.4888}) to ZnSe quantum wells ($\sim 15$ meV \cite{PhysRevA.106.L030402}). Applying this scaling factor ($\sim 1.9\times$) to the nitrogen acceptor baseline predicts a localization energy of 30 meV. Our observed value of $33.9 \pm 7.9$ meV is in excellent agreement with this prediction, confirming the emitter is probably a nitrogen acceptor.
Crucially, no extrinsic acceptor dopant source was introduced during the MBE growth process. This effectively rules out other common shallow acceptors, such as isoelectronic oxygen or lithium, which typically require specific precursors. Consequently, we attribute the acceptor population to residual nitrogen background in the vacuum chamber, which is a pretty common particle in other quantum emitter. 

%% file: sectionsSI/SectionI.tex
\section{Excited state lifetime}

\begin{figure}[h]
  \centering
  \includegraphics[width=448 px]{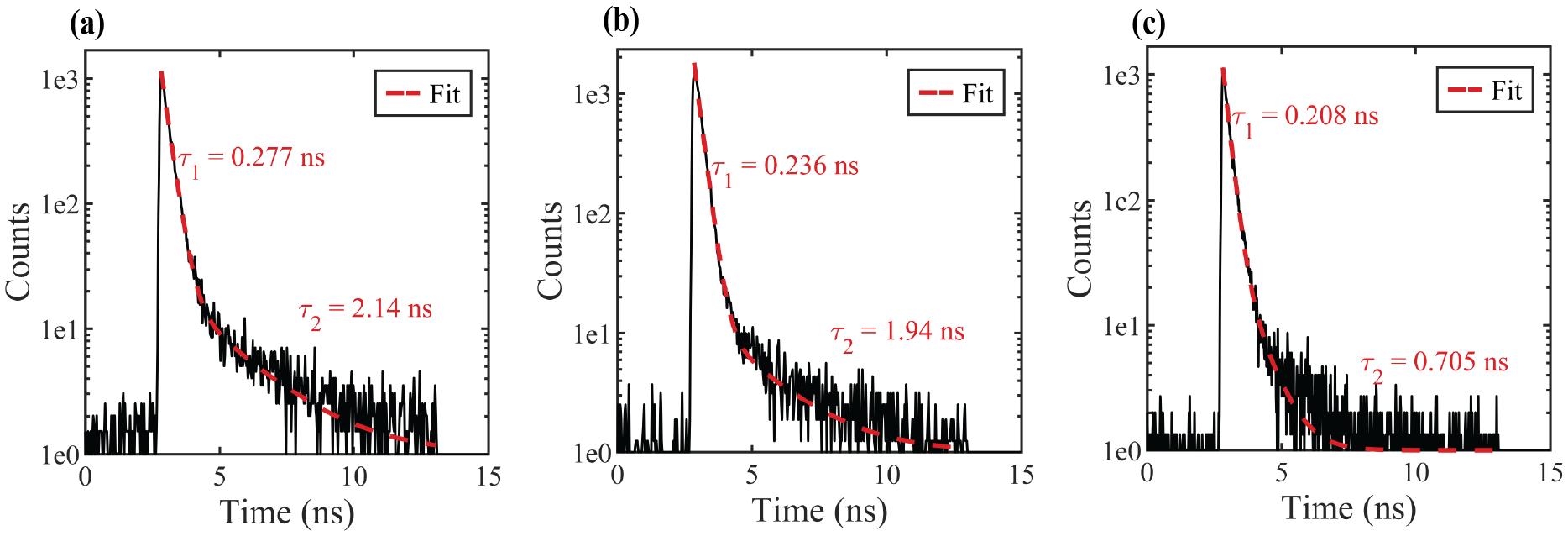}
  \caption{(a) (b) (c) Excited state lifetime measurement on three distinct acceptor sites.}
  \label{fig:S1}
\end{figure}

Fig. \ref{fig:S1} shows time-resolved measurements of the excited-state lifetime for three distinct acceptor-bound exciton emissions. Based on the total four extracted lifetimes (considering the one in main text), we report the average optical lifetime of $244 \pm 29$ ps for the individual nitrogen-bound exciton in ZnSe quantum well.

%% file: sectionsSI/Section0.tex
\section{RATE-EQUATION MODEL FOR OPTICAL CHARGE INJECTION}

Here we derive the rate-equation model used to analyze the photoluminescence excitation intensity as a function of above-band and resonant excitation powers, as presented in Fig.~3(c) of the main text. The model describes the population dynamics of an isolated acceptor impurity, which can exist in three distinct charge states.

\subsection{The Three-Level System}

The behavior of the acceptor is modeled using the three relevant states illustrated in Fig.~3(b) of the main text:
\begin{enumerate}
    \item $\Aneg$: The ionized acceptor state. In the n-doped ZnSe quantum well, this is the dark, equilibrium ground state. It has captured an electron from the donor-rich environment and cannot bind an exciton.
    \item $\Azero$: The neutral acceptor ground state. This state is created when the above-band excitation (ABE) pump generates a free hole, which is subsequently captured by an $\Aneg$ center. This state is required to form the bound exciton.
    \item $\AX$: The acceptor-bound exciton state. This is the excited state, formed when a $\Azero$ center captures a resonantly-created exciton (driven by the RE pump). The radiative decay from this state is the measured signal.
\end{enumerate}
We denote the steady-state populations of these levels as $N_-$, $N_0$, and $N_X$, respectively, with the constraint that $N_- + N_0 + N_X = 1$.

\subsection{Transition Rates}

The populations of these three states are governed by a few key processes. The rates for processes driven by optical power ($P$) are given by $R_i = k_i P_i$, where $k_i$ is a pumping coefficient (in units of W$^{-1}$s$^{-1}$) that depends on the absorption cross-section, spot size, and photon energy. The neutralization of the acceptor ($\Aneg \to \Azero$) is driven by two processes:
\begin{itemize}
    \item $\Rabe = k_{\text{ABE}} \Pabe$: The rate of neutralization ($\Aneg \to \Azero$), driven by the above-band (405 nm) pump. This process "supplies" the neutral ground state.
    \item $\Rleak = k_{\text{leak}} \Pre$: The rate of neutralization driven by the resonant (437 nm) pump. This leakage channel (e.g., via free-exciton creation and capture) also acts to supply the neutral ground state.
\end{itemize}
We define the total neutralization rate as $R_{\text{neut}} = \Rabe + \Rleak$. The other key processes are:
\begin{itemize}
    \item $\Rre = k_{\text{RE}} \Pre$: The rate of resonant excitation ($\Azero \to \AX$), driven by the resonant (437 nm) pump. 
    \item $\Grad$: The spontaneous radiative decay rate ($\AX \to \Azero$). This is a constant rate given by the inverse of the excited state lifetime, $\tau_1$.
    \item $\Gcap$: The non-resonant electron capture rate ($\Azero \to \Aneg$). This is a constant relaxation rate where a neutral acceptor captures an electron from the n-doped environment.
    \item $\Rion = k_{\text{ion}} \Pre$: The RE-induced ionization rate ($\Azero \to \Aneg$). This is a loss channel where the strong resonant laser itself provides sufficient energy to re-ionize the neutral acceptor.
\end{itemize}
The measured intensity $I$ (in counts/s) is the rate of photon emission ($\Grad N_X$) multiplied by the total collection and detection efficiency of the optical setup, $\eta$:
\begin{equation}
    I = \eta \cdot \Grad N_X \label{eq:intensity_definition}
\end{equation}

\subsection{Steady-State Rate Equations}

In the steady-state, the time derivatives of the populations are zero ($\frac{dN_i}{dt} = 0$). The balance equations for the three levels are:
\begin{align}
    \frac{dN_-}{dt} &= 0 = (\Gcap + \Rion)N_0 - R_{\text{neut}} N_- \label{eq:rate_Nminus}\\
    \frac{dN_X}{dt} &= 0 = \Rre N_0 - \Grad N_X \label{eq:rate_NX}\\
    \frac{dN_0}{dt} &= 0 = R_{\text{neut}} N_- + \Grad N_X - (\Rre + \Gcap + \Rion)N_0
\end{align}
We also have the conservation of population:
\begin{equation}
    N_- + N_0 + N_X = 1
    \label{eq:population_conservation}
\end{equation}
We only need to solve two of the first three equations along with the conservation equation.

\subsection{Solving for Intensity}

We aim to find the measured intensity $I$ as a function of $\Pabe$ (our x-axis) for a fixed $\Pre$ (our curves).

From Eq. \eqref{eq:rate_NX}, we can express $N_X$ in terms of $N_0$:
\begin{equation}
    N_X = \left( \frac{\Rre}{\Grad} \right) N_0 \label{eq:NX_solution}
\end{equation}
From Eq. \eqref{eq:rate_Nminus}, we can express $N_-$ in terms of $N_0$:
\begin{equation}
    N_- = \left( \frac{\Gcap + \Rion}{R_{\text{neut}}} \right) N_0
\end{equation}
Now, we substitute both $N_X$ and $N_-$ into the conservation equation, Eq. \eqref{eq:population_conservation}:
\begin{equation}
    \left( \frac{\Gcap + \Rion}{R_{\text{neut}}} \right) N_0 + N_0 + \left( \frac{\Rre}{\Grad} \right) N_0 = 1
\end{equation}
We solve for $N_0$:
\begin{equation}
    N_0 = \frac{1}{1 + \frac{\Rre}{\Grad} + \frac{\Gcap + \Rion}{R_{\text{neut}}}}
\end{equation}
Finally, we find $I$ by substituting this $N_0$ back into Eq.~\eqref{eq:NX_solution} and then into Eq.~\eqref{eq:intensity_definition}:
\begin{equation}
    I = \eta \cdot \Grad \cdot N_X = \eta \cdot \Grad \cdot \left( \frac{\Rre}{\Grad} \right) N_0
    \implies I = \frac{\eta \Rre}{1 + \frac{\Rre}{\Grad} + \frac{\Gcap + \Rion}{R_{\text{neut}}}} \label{eq:intensity_general}
\end{equation}
we substitute $R_{\text{neut}} = k_{\text{ABE}}\Pabe + k_{\text{leak}}\Pre$ into Eq.~\eqref{eq:intensity_general}:
\begin{equation}
    I(\Pabe) = \frac{\eta\ k_{\text{RE}} \Pre}{1 + \frac{k_{\text{RE}} \Pre}{\Grad} + \frac{\Gcap + k_{\text{ion}} \Pre}{k_{\text{ABE}}\Pabe + k_{\text{leak}}\Pre}}
    \label{eq:intensity_power_dependent}
\end{equation}
This equation describes the measured intensity as a function of all transition rates. 

\subsection{Fit Function and Analysis}
To fit the data in Fig.~3(c), we rearrange Eq.~\eqref{eq:intensity_power_dependent} into a standard generalized saturation form
\begin{equation}
    I(\Pabe) = \frac{\Isat \cdot \Pabe + I_0 \Psat}{\Pabe + \Psat}
\end{equation}
After some calculation steps, we can identify the key physical parameters. 
\begin{equation}
    \Isat = \eta \cdot \frac{k_{\text{RE}} \Pre}{1 + k_{\text{RE}} \Pre / \Grad}
    \label{eq:Isat}
\end{equation}
\begin{equation}
    \Psat = \frac{(1 + k_{\text{RE}}\Pre/\Grad)(k_{\text{leak}}\Pre) + (\Gcap + k_{\text{ion}}\Pre)}{k_{\text{ABE}}(1 + k_{\text{RE}}\Pre/\Grad)}
    \label{eq:Psat}
\end{equation}
\begin{equation}
    I_0 = I(\Pabe=0) = \frac{\eta (k_{\text{RE}}\Pre) (k_{\text{leak}}\Pre)}{(1 + k_{\text{RE}}\Pre/\Grad)(k_{\text{leak}}\Pre) + (\Gcap + k_{\text{ion}}\Pre)}
    \label{eq:I0}
\end{equation}

These equations perfectly describe the behavior in Fig.~3(c):
\begin{enumerate}
    \item \textbf{Saturation Intensity ($\Isat$):} The saturated plateau intensity, defined by Eq.~\eqref{eq:Isat}, is identical to the original model. It depends only on the resonant drive $\Rre$, and correctly increases as $\Pre$ increases.
    \item \textbf{Saturation Power ($\Psat$):} The effective saturation power of the above-band pump, defined by Eq.~\eqref{eq:Psat}, now has two terms. It depends on $\Rre$, $\Rion$, and now also $\Rleak$. Since all three of these rates increase with $\Pre$, the entire term $\Psat$ increases for stronger $\Pre$. This is seen clearly in Fig.~3(c) as the curves for higher $\Pre$ rise more slowly.
    \item \textbf{Y-Intercept Intensity ($I_0$):} The model now correctly predicts a non-zero y-intercept $I_0$, defined by Eq.~\eqref{eq:I0}, which is the signal generated by the $\Rleak$ supply rate alone. For the case of the emitter shown in the manuscript, this value is negligible, meaning that $k_{leak}$ almost zero. 
\end{enumerate}
This model confirms that $\Pabe$ acts to supply the ground state $\Azero$ against depletion, while $\Pre$ acts to both drive the $\AX$ transition and, at high powers, depopulate the $\Azero$ state via ionization.

\subsection{Experimental results}
In contrast to the theoretical prediction, in Fig. 3(c) we are not able to observe $I_0$ for different resonant levels. We attribute this to the lower contrast of signal to background noise at higher resonant power, which scales the background but the emitter stays dim in the absence of necessary above-band illumination. It could also emerge from $k_{\mathrm{leak}}\approx0$ for that emitter. 

\begin{figure}[h]
  \centering
  \includegraphics[width=400 px]{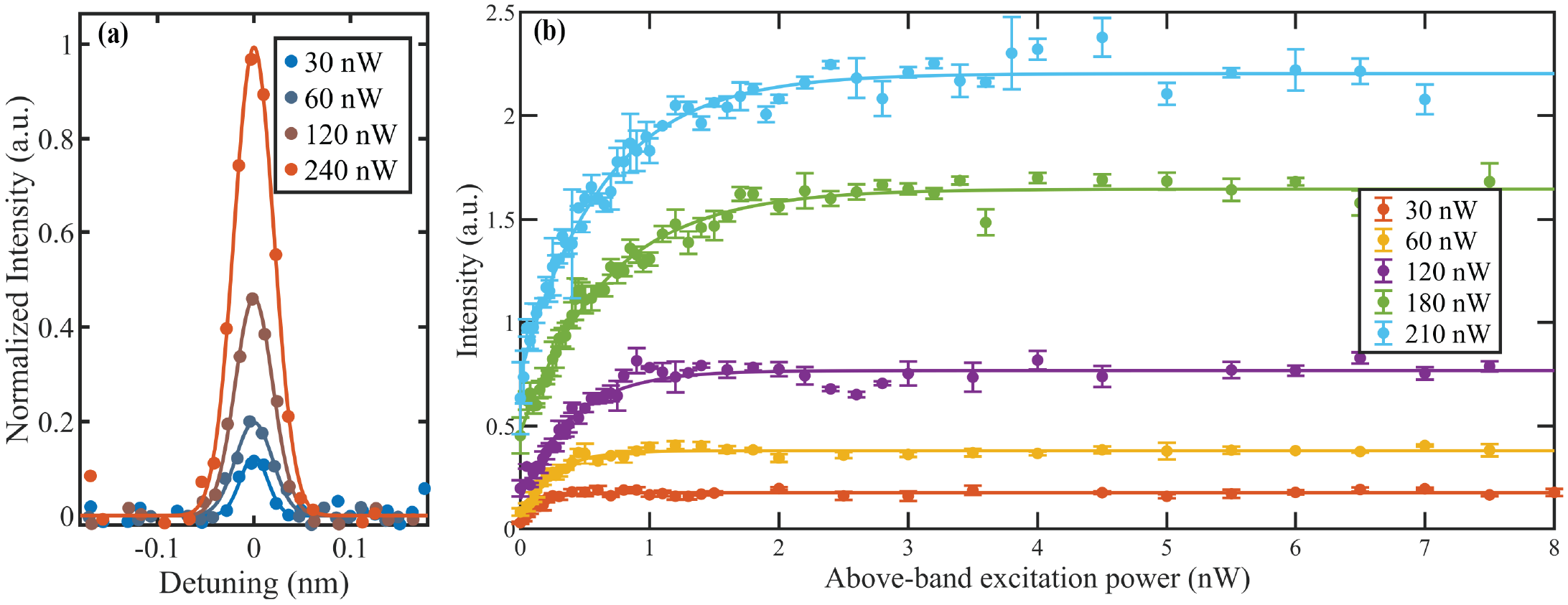}
  \caption{(a) Intensity of acceptor emission as a function of resonant laser power in the absence of above-band illumination. This shows that resonant laser can partially neutralize/activate the acceptor, although not as efficient as above-band. (b) Intensity of emission as a function of above-band power for different resonant powers. This indicates that above-band is still necessary to serve as the main source of neutralizing the acceptor.}
  \label{fig:S2}
\end{figure}

Here, we present a different emitter with improved background contrast, demonstrating the predicted partial saturation behavior and modulation of $I_0$ as a function of resonant power. Fig. ~\ref{fig:S2}(a) shows the intensity of emission in the absence of above-band for different resonant power levels. This indicates that the acceptor can be partially active even in the absence of above-band power. This observation confirms the theory at which resonant laser can partially neutralize the acceptor.  Fig. ~\ref{fig:S2}(b) shows the intensity of the emission as a function of above-band power for different resonant power levels. This indicates the main source of $\Aneg \to \Azero$ transition is above-band illumination. 

\clearpage